\documentclass[a4paper, 10pt]{article}
\usepackage{amsmath}
\usepackage{amsthm}
\usepackage{amssymb}
\usepackage{latexsym}
\usepackage{multirow}
\usepackage{dsfont}
\usepackage{booktabs}
\usepackage{graphicx}

\newtheorem{theorem}{Theorem}[section]
\newtheorem{definition}[theorem]{Definition}

\newtheorem{corollary}[theorem]{Corollary}

\newtheorem{lemma}[theorem]{Lemma}

\def\wt{\textrm{wt}}

\title{Complex Orthogonal Designs with Forbidden $2 \times 2$ Submatrices}
\author{Yuan Li and Haibin Kan}
\date{}

\begin{document}

\maketitle

\begin{abstract}
%\boldmath
Complex orthogonal designs (CODs) are used to construct space-time
block codes. COD $\mathcal{O}_z$ with parameter $[p, n, k]$ is a $p
\times n$ matrix, where nonzero entries are filled by $\pm z_i$ or
$\pm z^*_i$, $i = 1, 2, \ldots, k$, such that $\mathcal{O}^H_z
\mathcal{O}_z = (|z_1|^2+|z_2|^2+\ldots+|z_k|^2)I_{n \times n}$.
Define $\mathcal{O}_z$ a first type COD if and only if $\mathcal{O}_z$
does not contain submatrix $\begin{pmatrix} \pm z_j & 0
\\ \ 0 & \pm z^*_j
\end{pmatrix}$ or $\begin{pmatrix} \pm z^*_j & 0 \\ \ 0 & \pm z_j
\end{pmatrix}$. It is already known that, all CODs with
maximal rate, i.e., maximal $k/p$, are of the first type.

 In this paper, we determine all achievable parameters $[p, n, k]$
 of first type COD, as well as all their possible structures. The existence of parameters is proved
 by explicit-form constructions. New CODs with parameters
 $
 [p,n,k]=[\binom{n}{w-1}+\binom{n}{w+1}, n, \binom{n}{w}],
 $
 for $0 \le w \le n$, are constructed, which demonstrate the possibility of sacrificing code rate to reduce decoding delay.
    It's worth
 mentioning that all maximal rate, minimal delay CODs are contained in our constructions, and their uniqueness under equivalence operation is proved.
\end{abstract}

\section{Introduction}
Space-time block codes have been widely investigated for wireless
communication systems with multiple transmit and receive antennas.
Since the pioneering work by Alamouti \cite{Ala98} in 1998, and the
work by Tarokh et al. \cite{TJC99}, \cite{TJC00}, orthogonal designs
have become an effective technique for the design of space-time
block codes (STBC). The importance of this class of codes comes from
the fact that they achieve full diversity and have the fast
maximum-likelihood (ML) decoding.

A complex orthogonal design (COD) $\mathcal{O}_z[p, n, k]$ is an $p
\times n$ matrix, and each entry is filled by $\pm z_i$ or $\pm
z^*_i$, $i=1,2,\ldots, k$, such that $\mathcal{O}^H_z \mathcal{O}_z
= \sum_{i=1}^{n}|z_i|^2 I_n$, where $H$ is the Hermitian transpose
and $I_n$ is the $n \times n$ identity matrix. Under this
definition, the designs are said to be combinatorial, in the sense
that there is no linear processing in each entry. When linear
combination of variables are allowed, we call it generalized complex
orthogonal design (GCOD).

Code rate $k/p$ and decoding delay $p$ are the two most important
criteria of complex orthogonal space-time block codes. One important
problem is, given $n$, determine the tight upper bound of code rate,
which is called maximal rate problem. Another is, given $n$,
determine the tight lower bound of decoding delay $p$ when code rate
$k/p$ reaches the maximal, which is called minimal delay problem.

For combinatorial CODs, where linear combination is not allowed,
Liang determined for a COD with $n=2m$ or $2m-1$, the maximal
possible rate is $\frac{m+1}{2m}$ \cite{Lia03}. Liang gave an
algorithm in \cite{Lia03} to generate such CODs with rate
$\frac{m+1}{2m}$, which shows that this bound is tight. In
\cite{LSG10}, Yuan et al. simplifies Liang's proof on the upper
bound of code rate slightly. The minimal delay problem are solved by
Adams et al. In \cite{AKP07}, lower bound ${2m \choose m-1}$ of
decoding delay is proved for any $n=2m$ or $2m-1$. In \cite{AKM10},
Adams et al. prove that when $n \equiv 2 \pmod 4$, decoding delay
$p$ is lowered bound by $2{2m \choose m-1}$.

Besides some scattered constructions for relatively small number of
antennas $n$ \cite{WXS03}, \cite{TJC99}, \cite{SX02}, several
general methods to construct complex orthogonal designs have been
proposed. Liang's algorithmic construction in \cite{Lia03} achieves
the maximal rate for all $n$, acheives the minimal delay when $n
\equiv 1, 2, 3 \pmod 4$. But when $n \equiv 0 \pmod 4$, the delay is
twice of the minimal delay. In \cite{SXL04}, a different algorithmic
method to generate complex orthogonal is proposed, which has the
same code rate and decoding delay as Liang's construction. In
\cite{LFX05}, a closed-form iterative construction of complex
orthogonal designs was proposed, which achieves both the maximal
rate and minimal delay.

For GCOD, which allows linear combination in each entry, little is
known about the rate and delay. In \cite{LX03}, they proved that
there does not exist rate 1 GCOD when $n \ge 3$. In \cite{WX03},
Wang and Xia proved an upper bound $4/5$ of the code rate for GCODs
without equal weight condition, and an upper bound $3/4$ with equal
weight condition when $n \ge 3$. And this result is the best as far
as we know.

The unfortunate property of COD is that for $n=2m$ or $2m-1$
transmit antennas, the codes with maximal rate $(m+1)/(2m)$ has
minimal decoding delay ${2m \choose m-1}$(with exception $n=2\pmod4$
where it is $2{2m \choose m-1}$). For example, when $n=14$, the
minimal delay for a code with maximal rate is 6006! Therefore, it's
meaningful to construct CODs with smaller decoding delay by
sacrificing code rate and investigate the tradeoff between code rate
and decoding delay. For example, in \cite{ADK11}, Adams et al.
considered a class of CODs with rate $\frac{1}{2}$ and proved a
lower bound on delay.

In this paper, by restricting to a specific type of CODs which
contains no submatrices  $\begin{pmatrix} \pm z_j & 0
\\ \ 0 & \pm z^*_j
\end{pmatrix}$ or $\begin{pmatrix} \pm z^*_j & 0 \\ \ 0 & \pm z_j
\end{pmatrix}$, which are called first type CODs in the
paper, we consider the most general problem that determining what
parameters $[p, n, k]$ are achievable. Not only all achievable
parameters are determined, but also all their possible structures
are also proved. It should be noticed that all CODs with maximal
rate are of first type, and thus it is  not a very strict restriction.

The organization of our paper is as follows. In section 2, we
introduce the notions which will be used. In section 3, we review
some basic definition and some known results about CODs. In section
4, we present our explicit-form constructions. In section 5, we
prove our constructions in section 4 consist of all first type CODs, up
to equivalence operation and simple catenation operation. In section
5, we give out the conclusions.

\section{Notations}

In this section, we introduce some basic notions, which will be used
in the sequel.

$\mathds{C}$ denotes the field of complex numbers, $\mathds{R}$ the
field of real numbers and $\mathbb{F}_2$ the field with two
elements. Adding over $\mathbb{F}_2$ is denoted by $\oplus$ to avoid
ambiguity. All vectors are assumed to be column vectors. For any
field $\mathds{F}$, denoted by $\mathds{F}^n$ and $M_{m \times
n}(\mathds{F})$ the set of all $n$-dimensional vectors in
$\mathds{F}$ and the set of all $m \times n$ matrices in
$\mathds{F}$, respectively. For any vector $x \in \mathds{F}^n$,
denote by $x^T$ the transpose of $x$. For any matrix $A \in M_{m
\times n}(\mathds{C})$, denote by $A^T$ the transpose of $A$ and by
$A^H$ the conjugate transpose of $A$. Denote by
\begin{displaymath}
A(i_1,i_2,\ldots,i_p;j_1,j_2,\ldots,j_q) \text{ and } A(s_1,\ldots,
s_2; t_1,\ldots,t_2)
\end{displaymath}
the submatrix consisting of $i_1^{\text{th}}$, $i_2^{\text{th}}$,
$\ldots$, $i_p^{\text{th}}$ rows and the $j_1^{\text{th}}$,
$j_2^{\text{th}}$, $\ldots$, $j_q^{\text{th}}$ columns of $A$, and
the submatrix consisting of the $s_1^{\text{th}}$,
$(s_1+1)^{\text{th}}$, \ldots, $s_2^{\text{th}}$ rows and the
$t_1^{\text{th}}$, $(t_1+1)^{\text{th}}$, \ldots, $t_2^{\text{th}}$
columns of $A$, where $s_1 < s_2$ and $t_1 < t_2$, respectively. We
use $A(i,j)$ for the $(i,j)$ element of the matrix $A$. In this
paper, rows and variables are often indexed by vectors in
$\mathbb{F}^2_n$.

For convenience, let $e_i \in \mathbb{F}^n_2$ be the vector with
$i^{\text{th}}$ bit occupied by $1$ and the others $0$, i.e.,
$
e_i = (\underbrace{0, \ldots, 0}_{i-1}, 1, \underbrace{0, \ldots,
0}_{n-i})
$
and let $e = e_1 \oplus e_2 \oplus \ldots \oplus e_n$, i.e.,
$
e = (1, 1, \ldots, 1)_2.
$

The weight of a vector in $\mathbb{F}^n_2$ is defined as the number
of ones in $n$ bits, i.e., $\wt(\alpha) = \sum_{i=1}^{n} \alpha(i)$.
Furthermore, $\wt_{s, t}(\alpha)$ is defined as the sum of
$s^{\text{th}}$ bit to $t^{\text{th}}$ bit, i.e.,
\begin{displaymath}
\wt_{s, t}(\alpha) = \alpha(s) + \alpha(s+1) + \cdots + \alpha(t) =
\sum_{i = s}^{t}{\alpha(i)}.
\end{displaymath}

In abuse of notation, we denote by $z[j]$ the complex variable
$z_j$, up to negation and conjugation, i.e., $z[j] \in \{ z_j, -z_j,
z^*_j, -z^*_j\}.$ Note that the same notation $z[j]$ may represent
different elements in the same paragraph.

%\vspace{0.4cm}

\section{Definitions and Some Known Results}

\begin{definition}
A $[p, n, k]$ complex orthogonal design $\mathcal{O}_z$ is a $p
\times n$ rectangular matrix whose nonzero entries are
\begin{displaymath}
z_1, z_2, \ldots, z_{k}, -z_1, -z_2, \ldots, -z_{k}
\end{displaymath}
or their conjugates
\begin{displaymath}
z_1^*, z_2^*, \ldots, z_{k}^*, -z_1^*, -z_2^*, \ldots, -z_{k}^*,
\end{displaymath}
where $z_1, z_2, \ldots, z_{k}$ are indeterminates over
 $\mathds{C}$, such that
\begin{displaymath}
\mathcal{O}_z^H\mathcal{O}_z = (|z_1|^2+|z_2|^2+\cdots+|z_{k}|^2)
I_{n \times n}.
\end{displaymath}
$k/p$ is called the code rate of $\mathcal{O}_z$, and $p$ is called
the decoding delay of $\mathcal{O}_z$.
\end{definition}

A matrix is called an Alamouti $2 \times 2$ if it matches the
following form
\begin{equation}
\label{equ:Alamouti}
\begin{pmatrix}
z_i & z_j \\
-z_j^* & z_i^*
\end{pmatrix},
\end{equation}
up to negation or conjugation of $z_i$ or $z_j$. We say two rows
share an Alamout $2 \times 2$ if and only if the intersection of the
two rows and some two columns form an Alamouti $2 \times 2$.

\begin{definition}
The equivalence operations performed on any COD are defined as
follows.
\begin{itemize}
\item[] 1) Rearrange the order the rows(``row permutation'').
\item[] 2) Rearrange the order the columns (``column permutation'').
\item[] 3) Conjugate all instances of certain variable (``instance
conjugation'').
\item[] 4) Negate all instances of certain variable (``instance
negation'').
\item[] 5) Change the index of all instances of certain variable (``instance
renaming'').
\item[] 6) Multiply any row by $-1$, (``row negation'').
\item[] 7) Multiply any column by $-1$, (``column negation'').
\end{itemize}
\end{definition}

It's not difficult to verify that, given a COD
$\mathcal{O}_z[p,n,k]$, after arbitrary equivalence operations, we
will obtain another COD $\mathcal{O}'_z[p,n,k]$. And we say COD
$\mathcal{O}_z$ and $\mathcal{O}'_z$ are the same under equivalence
operations.

Following the definition in \cite{Lia03}, define an $(n_1,
n_2)$-$\mathcal{B}_j$ form by
\begin{eqnarray}
\mathcal{B}_j & = & \begin{pmatrix} z_j I_{n_1} & \mathcal{M}_1 \\
-\mathcal{M}^H_1 & z^*_j I_{n_2}
\end{pmatrix} \nonumber\\
& = &
       \left(
       \begin{array}{c|c}
       \begin{array}{cccc}
       z_j&0&\cdots&0\\
       0&z_j&\cdots&0\\
       \vdots&\vdots&\ddots&\vdots\\
       0&0&\cdots&z_j
       \end{array}&\mathcal{M}_j\\
       \hline
       -\mathcal{M}_j^H&\begin{array}{cccc}
       z_j^*&0&\cdots&0\\
       0&z_j^*&\cdots&0\\
       \vdots&\vdots&\ddots&\vdots\\
       0&0&\cdots&z_j^*
       \end{array}
       \end{array}
       \right), \label{equ:Bj_form}
\end{eqnarray}
where $n_1 + n_2 = n$. And we call it $\mathcal{B}_j$ form for
short.

\begin{definition} \cite{AKP07}
We say COD $\mathcal{O}_z$ is in $\mathcal{B}_j$ form if the
submatrix $\mathcal{B}_j$ can be created from $\mathcal{O}_z$
through equivalence operations except for column permutation.
Equivalently, $\mathcal{O}_z$ is in $\mathcal{B}_j$ form if every
row of $\mathcal{B}_j$ appears within the rows of $\mathcal{O}_z$,
up to possible negations or conjugations of all instances of $z_i$ and possible
factors of $-1$.
\end{definition}

 It is proved that
\cite{AKP07} that COD $\mathcal{O}_z$ is in some $\mathcal{B}_j$
form if and only if one row in $\mathcal{O}_z$ matches one row of
$\mathcal{B}_j$ up to signs and conjugations.

In \cite{Lia03}, Liang proved the upper bound $\frac{m+1}{2m}$ of
code rate $\frac{k}{p}$ for any $n=2m$ or $2m-1$, and obtained the necessary and
sufficient condition to reach the maximal rate.

\begin{theorem}
\label{thm:max_rate} Let $n=2m$ or $2m-1$. The rate of COD
$\mathcal{O}_z[p,n,k]$ is upper bounded by $\frac{m+1}{2m}$, i.e.,
$\frac{k}{p} \le \frac{m+1}{2m}$.

This bound is achieved if and only if for all $j=1,2,\ldots, k$,
$\mathcal{B}_j$ is an $(m,m-1)$-$\mathcal{B}_j$ or
$(m-1,m)$-$\mathcal{B}_j$ form and there are no zero entries in
$\mathcal{M}_j$, when $n=2m-1$; $\mathcal{B}_j$ is an
$(m,m)$-$\mathcal{B}_j$ form and there are no zero entries in
$\mathcal{M}_j$, when $n=2m$.
\end{theorem}

The lower bound on the decoding delay when code rate reaches the maximal
is completely solved by Adams et al. in \cite{AKP07} and
\cite{AKM10}.

\begin{theorem} Let $n=2m$ or $2m-1$. For COD
$\mathcal{O}_z[p,n,k]$, if the rate reaches the maximal, i.e.,
$\frac{k}{p} = \frac{m+1}{2m}$, the delay $p$ is lower
bounded by ${2m \choose m-1}$ when $n \equiv 0,1,3 \pmod 4$; by
$2{2m \choose m-1}$ when $n \equiv 2 \pmod 4$.
\end{theorem}

The technique in proving the lower bound ${2m \choose m-1}$ is the
observation and definition of zero pattern, which is a vector in
$\mathbb{F}^n_2$ defined with respect to one row where the
$i^{\text{th}}$ bit is $0$ if and only if the element on column $i$
is $0$. For example, when
\begin{equation}
\mathcal{O}_z=
\begin{pmatrix}
z_1 & z_2 & z_3 \\
-z^*_2 & z^*_1 & 0 \\
-z^*_3 & 0 & z^*_1 \\
0 & z^*_3 & -z^*_2
\end{pmatrix},
\label{equ:COD433}
\end{equation}
the first row has zero pattern $(1,1,1)$, the second $(1,1,0)$, the
third $(1,0,1)$, the fourth $(0,1,1)$.

Next, we propose some new definitions.

\begin{definition} COD $\mathcal{O}_z[p,n,k]$ is a first type COD if it does not contain submatrix
$$
\begin{pmatrix} \pm z_j & 0 \\ 0 & \pm z^*_j
\end{pmatrix} { \text{ or }}
\begin{pmatrix} \pm z^*_j & 0 \\ 0 & \pm z_j
\end{pmatrix},
$$
where $1 \le j \le k$.
\end{definition}

In other words, COD $\mathcal{O}_z[p,n,k]$ is of first type if for all
$1 \le j \le k$, there is no zero entry in $\mathcal{M}_j$ of its
$\mathcal{B}_j$ form. By Theorem \ref{thm:max_rate}, we can see all
maximal-rate CODs are in the first type.

\begin{definition} COD $\mathcal{O}_z[p,n,k]$ is
called atomic if and only if there does not exist a COD which is a
submatrix of $\mathcal{O}_z$ consisting of some (not all)
rows of $\mathcal{O}_z$.

Formally, $\mathcal{O}_z[p,n,k]$ is atomic if and only if for any
integers $1\le q \le p-1, 1 \le i_1 < i_2 < \cdots \ i_{q} \le p$,
$\mathcal{O}_z(i_1, i_2, \ldots, i_q; 1,\ldots ,n)$ is a not COD.
Otherwise, $\mathcal{O}_z[p,n,k]$ is called non-atomic.
\end{definition}

For an atomic COD $\mathcal{O}_z[p, n, k]$, given any $1 \le s, t\le
k$, there exist $j_1=s, j_2, \ldots, j_{m-1}, j_m=t$ such that
$\mathcal{B}_{j_1}$ and $\mathcal{B}_{j_2}$ share some common rows,
$\mathcal{B}_{j_2}$ and $\mathcal{B}_{j_3}$ share some common rows,
..., $\mathcal{B}_{j_{m-1}}$ and $\mathcal{B}_{j_m}$ share some
common rows. This condition is also sufficient for a COD to be
atomic.

For COD $\mathcal{O}_z$, assume one row is in some atomic COD
$\mathcal{O}'_z$ which consists of some rows of $\mathcal{O}_z$. If
one variable is in $\mathcal{O}'_z$, then all rows containing this
variable is in $\mathcal{O}'_z$. Repeat this procedure until no more
rows are added. Finally, atomic COD $\mathcal{O}'_z$ is obtained. By
the above algorithm, we can see COD $\mathcal{O}_z$ can be
decomposed into atomic ones in a unique way.

For example, let $\mathcal{O}_z$ consists of the first two columns of \eqref{equ:COD433}, i.e.,
\begin{equation}
\mathcal{O}_z=
\begin{pmatrix}
z_1 & z_2\\
-z^*_2 & z^*_1 \\
-z^*_3 & 0\\
0 & z^*_3
\end{pmatrix}.
\end{equation}
Then $\mathcal{O}_z$ can be decomposed into two atomic ones
$$
\begin{pmatrix} z_1 & z_2 \\ -z^*_2 & z_1^*
\end{pmatrix} { \text{ and }}
\begin{pmatrix} -z^*_3 & 0 \\ 0 & z^*_3
\end{pmatrix}.
$$

On the contrary to the decomposition of COD, given two (or more) CODs with
parameters $\mathcal{O}_1[p_1, n, k_1]$ and $\mathcal{O}_2[p_2, n,
k_2]$, we can construct a new COD with parameter $[p_1 + p_2, n,
k_1+k_2]$ by simply catenating them, i.e., $\begin{pmatrix}
\mathcal{O}_1
\\ \mathcal{O}_2 \end{pmatrix}$, and renaming certain variables of $\mathcal{O}_1$ and $\mathcal{O}_2$
to avoid conflicts if necessary. We call it catenation operation.

\section{Explicit-form Constructions}

In this section, we present explicit-form constructions of first type
CODs. The basic idea is first to construct a basic COD with rate
$1/2$ and parameters $[2^{n+1}, n, 2^n]$, which are based on
combinatorial methods by using vectors in $\mathbb{F}_2^{n+1}$.
Then, by choosing submatrices from the basic COD, we obtain CODs
with parameters
$$
[p, n, k] = [{n \choose w-1}+{n \choose w+1}, n, {n \choose w}],
$$
where $-1 \le w \le n+1$. Note that, when $n \not\equiv 0 \pmod 4$, all maximal-rate, minimal-delay
CODs are contained in the above constructions.

Next, we consider $n \equiv 0 \pmod 4$. By padding an extra column
on our basic COD, we obtain COD with parameter $[2^n, n, 2^{n-1}]$.
Again, by choosing submatrices from the basic COD, we obtain CODs
with parameters $[{n \choose n/2+1},n,{n-1 \choose n/2-1}]$, which
are optimal.

\begin{theorem}
\label{thm:Gn} Let $\mathcal{G}_n$ be $2^{n+1} \times n$ matrix,
where rows are indexed by vectors in $\mathbb{F}_2^{n+1}$ and
columns are indexed by $1, 2, \ldots, n$. For all $\alpha \in
\mathbb{F}_2^{n+1}, 1 \le i \le n$,
\begin{itemize}
\item if $\alpha(i) = 0$, then $\mathcal{G}_n(\alpha, i) = 0$,
\item if $\alpha(i) = 1$ and $\alpha(n+1)=0$, then $\mathcal{G}_n(\alpha, i) = (-1)^{\theta(\alpha, i)}z_{\varphi(\alpha,
i)}$,
\item if $\alpha(i) = 1$ and $\alpha(n+1)=1$, then $\mathcal{G}_n(\alpha, i) = (-1)^{\theta(\alpha, i)}z^*_{\varphi(\alpha,
i)}$,
\end{itemize}
where
\begin{equation}
\label{equ:def_theta} \theta(\alpha, i) =
\begin{cases}
wt_{i, n+1}(\alpha) + \frac{i}{2}, & \text{if $i$ is even}, \\
wt_{i, n+1}(\alpha) + \frac{i-1}{2} + \alpha(n+1), & \text{if
$i$ is odd},
\end{cases}
\end{equation}
and
\begin{eqnarray}
\varphi(\alpha, i) & = & \alpha \oplus \alpha(n+1)e \oplus e_i \nonumber \\
 & = & (\alpha(1) \oplus \alpha(n+1), \ldots, \alpha(i)
\oplus \alpha(n+1) \oplus 1 , \ldots, \nonumber\\
& & \alpha(n+1) \oplus \alpha(n+1)). \nonumber
\end{eqnarray}
Then $\mathcal{G}_n$ is a COD with parameter $[2^{n+1}, n, 2^n]$.
\end{theorem}
\begin{proof} It is sufficient to
prove 1) every variable, up to negation or conjugation, appears
exactly once in each column; 2) any two different columns are
orthogonal.

Since for fixed $i$, $\varphi(\alpha, i)$ takes nonzero values on
$2^{n}$ different vectors $\alpha \in \mathbb{F}^{n+1}_2,
\alpha(i)=1$. To prove 1), we only need to show $\varphi$ is a
surjective, i.e. $\alpha \not= \beta \Rightarrow \varphi(\alpha, i)
\neq \varphi(\beta, i)$. Suppose to the contrary that there exists
$\alpha$ and $\beta$ where $\alpha \neq \beta$, $\alpha(i) =
\beta(i) = 1$ and $\varphi(\alpha, i) = \varphi(\beta, i)$.
Expanding $\varphi(\alpha, i) = \varphi(\beta, i)$ by definition, we
have
\begin{equation}
\alpha \oplus \alpha(n+1)e \oplus e_i = \beta \oplus \beta(n+1)e
\oplus e_i, \nonumber
\end{equation}
which is equivalent to
\begin{equation} \alpha \oplus \beta = (\alpha(n+1) \oplus \beta(n+1))e. \nonumber
\end{equation}
If $\alpha(n+1) = \beta(n+1)$, then $\alpha \oplus \beta =
(\alpha(n+1) \oplus \beta(n+1))e = 0\Rightarrow \alpha = \beta$,
which is contradicted with $\alpha \neq \beta$. If $\alpha(n+1) \neq
\beta(n+1)$, then $\alpha \oplus \beta = (\alpha(n+1) \oplus
\beta(n+1))e = e$, which is contradicted with $\alpha(i) = \beta(i)
= 1$.

To prove any two different columns are orthogonal, it is sufficient
to show that, every pair of nonzero entries in the same row are in
an Alamouti $2 \times 2$.

Let columns $1 \leq i < j \leq n$ and $\alpha \in
\mathbb{F}_2^{n+1}$ be any row, satisfying $\alpha(i) = \alpha(j) =
1$. Let $\gamma = \varphi(\alpha, i)$, $\delta = \varphi(\alpha,
j)$. Since every variable appears exactly once in each column, we
assume $z[\delta]$ appears in the $\beta^{\text{th}}$ row in $i^{\text{th}}$ column,
i.e., $\varphi(\beta, i) = \gamma$.

By the assumption that $z[\delta]$ appears in $\mathcal{G}_n(\beta,
i)$, we have $\varphi(\beta, i) = \varphi(\alpha, j)$, i.e.,
\begin{equation}
\alpha \oplus \alpha(n+1)e \oplus e_j = \beta \oplus \beta(n+1)e
\oplus e_i, \nonumber
\end{equation}
which implies
\begin{eqnarray}
\beta & = & \alpha \oplus \alpha(n+1)e \oplus e_{j} \oplus
\beta(n+1)e
\oplus e(i) \nonumber \\
& = & \alpha \oplus (\alpha(n+1) \oplus \beta(n+1))e \oplus e_i
\oplus e_{j}. \label{equ:s'_s}
\end{eqnarray}
Noting that $\varphi$ takes nonzero value on $(\alpha, i)$,
$(\alpha, j)$ and $(\beta, i)$, we have $\alpha(i) = \alpha(j) =
\beta(j) = 1$. Considering $i^{\text{th}}$ value in equality
\eqref{equ:s'_s}, we conclude $\alpha(n+1) \oplus \beta(n+1) = 1$.
Thus,
\begin{equation}
\label{equ:sum_s_s'} \alpha \oplus \beta = e \oplus e_i \oplus e_j.
\end{equation}
Taking $\beta=\alpha\oplus e\oplus e_i \oplus e_j$ into $\varphi(\beta, j)$, we have
\begin{eqnarray*}
\varphi(\beta, j) & = & \beta \oplus \beta(n+1)e \oplus e_{j} \\
& = & \alpha \oplus \alpha(n+1)e \oplus e_{j} \oplus \beta(n+1)e
\oplus e_i \oplus
\beta(n+1)e \oplus e_{j}\\
& = &\alpha \oplus \alpha(n+1)e \oplus e_{i}\\
& = & \varphi(\alpha, i).
\end{eqnarray*}
Therefore, submatrix $\mathcal{G}_n(\alpha, \beta; i, j)$ could be
written in either of the two following forms
\begin{equation}
\begin{pmatrix}
(-1)^{\theta(\alpha,i)}z_\gamma & (-1)^{\theta(\alpha, j)} z_\delta \\
(-1)^{\theta(\beta, i)} z_\delta^* & (-1)^{\theta(\beta, j)}
z_\gamma^*
\end{pmatrix} \nonumber
\end{equation}
or
\begin{equation}
\begin{pmatrix}
(-1)^{\theta(\alpha, i)}z_\gamma^* & (-1)^{\theta(\alpha, j)} z_\delta^* \\
(-1)^{\theta(\beta, i)} z_\delta & (-1)^{\theta(\beta, j)} z_\gamma
\end{pmatrix}. \nonumber
\end{equation}

Now we calculate $\theta(\alpha, i) + \theta(\alpha, j) +
\theta(\beta, i) + \theta(\beta, j)$ to check whether it is an Alamouti $2 \times 2$. First, let's
calculate $\theta(\alpha, i) + \theta(\beta, i)$ by \eqref{equ:def_theta}. When $i$ is even, $\theta(\alpha, i) + 
\theta(\beta, i) = \wt_{i, n+1} + \frac{i}{2} + \wt_{i, n+1}(\beta) + \frac{i}{2} \equiv \wt_{i, n+1}(\alpha \oplus \beta) + i
\pmod 2$; When $i$ is odd, $\theta(\alpha, i) + 
\theta(\beta, i) = \wt_{i, n+1} + \frac{i-1}{2} + \alpha(n+1) +  \wt_{i, n+1}(\beta) + \frac{i-1}{2} + \beta(n+1) \equiv \wt_{i, n+1}(\alpha \oplus \beta) + i \pmod 2$. Therefore, we have
\begin{equation}
\theta(\alpha, i) + 
\theta(\beta, i) = \wt_{i, n+1}(\alpha \oplus \beta) + i \pmod 2
\end{equation}
always holds.

Then,
\begin{eqnarray}
& & \theta(\alpha, i) + \theta(\beta, i)+\theta(\alpha, j) +
\theta(\beta, j) \nonumber\\
& \equiv & \wt_{i,2m}(\alpha \oplus \beta) + i + \wt_{j,2m}(\alpha
\oplus \beta) + j \nonumber\\
& \equiv & \wt_{i, j-1}(\alpha \oplus \beta) + i + j \nonumber\\
& \equiv & j-i-1+i+j \nonumber\\
& \equiv & 1 \pmod 2. \nonumber
\end{eqnarray}
In the last second step, $\wt_{i, j-1}(\alpha \oplus \beta) = j-i-1$
is true because $\alpha \oplus \beta = e \oplus e_i \oplus e_j$.

Therefore,
\begin{displaymath}
(-1)^{\theta(\alpha,i)}z_\gamma^*(-1)^{\theta(\alpha, j)}z_\delta +
(-1)^{\theta(\beta, i)}z_\delta(-1)^{\theta(\beta, j)}z_\gamma^* = 0
\end{displaymath}
holds and the submatrix $\mathcal{G}_n(\alpha, \beta; i, j)$ is an
Alamouti $2 \times 2$, which implies column $i$ and column $j$ are
orthogonal.
\end{proof}

By taking out some submatrices form $\mathcal{G}_n$, we can get a
series of atomic first type CODs.

\begin{theorem}
\label{Thm:Gnm} Given $n$, for arbitrary integer $-1 \le w \le n+1$,
let
$$\mathcal{G}^w_n = \mathcal{G}_n(\alpha_1, \ldots,
\alpha_{\binom{n}{w+1}}, \beta_1, \ldots, \beta_{\binom{n}{n-w+1}};
1,\ldots,n),$$ where $\alpha_i$ are all vectors in
$\mathbb{F}^{n+1}_2$ with weight $w+1$ and the $(n+1)^{\text{th}}$
bit $0$, $\beta_i$ are all vectors in $\mathbb{F}^{n+1}_2$ with
weight $n-w+2$ and the $(n+1)^{\text{th}}$ bit $1$. Then
$\mathcal{G}^w_n$ is a COD with parameter $[{n \choose w+1}+{n
\choose w-1}, n, {n \choose w}]$.
\end{theorem}
\begin{proof}
Since $\mathcal{G}_n^w$ is a submatrix of the orthogonal design
$\mathcal{G}_{n}$, it's sufficient to prove that if some variable
exists on one column of $\mathcal{G}^w_n$ then it exists on every
column of $\mathcal{G}^w_n$. We will show that all variables with
subscript weight $w$ exist on each column of $\mathcal{G}^w_n$.

For any $\alpha \in \mathbb{F}_2^{n+1}$ such that $\alpha(n+1)=0,
\alpha(i) = 1$ for some $1 \le i \le n$, as $\wt(\varphi(\alpha, i))
= \wt(\alpha \oplus e_i) = \wt(\alpha) - 1$, then
$\wt(\varphi(\alpha, i)) = w$ if and only if $\wt(\alpha) = w+1$.

For any $\alpha \in \mathbb{F}_2^{n+1}$ such that $\alpha(n+1)=1$,
and $\alpha(i) = 1$ for some $1 \leq i \leq n$, as
$\wt(\varphi(\alpha, i)) = \wt(\alpha \oplus e_i \oplus e) =
n+2-\wt(\alpha)$, then $\wt(\varphi(\alpha, i)) = w$ if and only if
$\wt(\alpha) = n-w+2$.

Finally, there are ${n \choose w+1} + {n \choose n-w+1}={n \choose
w+1} + {n \choose w-1}$ rows taken and ${n \choose w}$ different
variables in it.
\end{proof}

Notice that, in the above constructions, $\mathcal{G}^{-1}_n=(0, 0, \ldots, 0)$ is a trivial COD with rate $0$
and delay $1$.

For fixed number of antennas $n$, the code rate
$$\frac{{n \choose
w}}{{n \choose w+1}+{n \choose w-1}} = \left( \frac{n-w}{w+1} +
\frac{w}{n-w+1}\right)^{-1}$$ is an increasing function of $w$ when
$-1 \leq w \leq \lfloor \frac{n}{2} \rfloor$, as well as the
decoding delay ${n \choose w}$. Since the decoding delay ${n \choose
w}$ grows very fast when $w$ is increasing, the sacrifice in rate
might be worth the trade-off for a smaller decoding delay in
practice.

For example, let $n = 14$, $w = 0, 1, \ldots, 7$ respectively, we
obtain codes with the parameters with rate decreasing and
delay increasing in Figure 1.

\begin{figure}
\includegraphics[scale = 0.5]{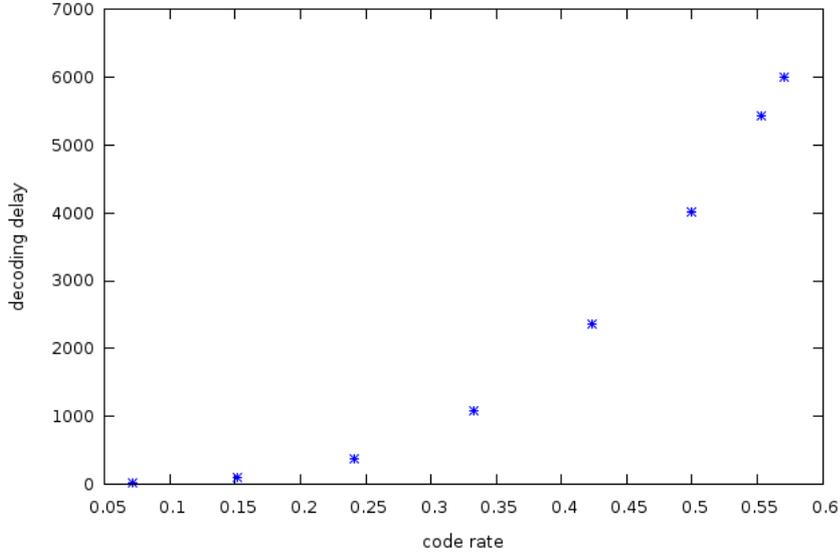}
\caption{Code rate and delay for $n = 14$}
\end{figure}
%\begin{table}[htbp]
%\centering
% \caption{\label{tab:test}Code rate and delay for $n = 14$}
% \begin{tabular}{cccc}
%  \toprule
%  $w$ & $k$ & rate = $k / p$ & delay = $p$\\
%  \midrule
%0 & 1 & 0.071 & 14\\
%1 & 14 & 0.152 & 92\\
%2 & 91 & 0.241 & 378\\
%3 & 364 & 0.333 & 1092\\
%4 & 1001 & 0.423 & 2366\\
%5 & 2002 & 0.500 & 4004\\
%6 & 3003 & 0.553 & 5434\\
%7 & 3432 & 0.571 & 6006\\
%  \bottomrule
% \end{tabular}
%\end{table}

Like the  Alamouti $2 \times 2$ in \cite{Ala98}, certain CODs enjoy
a property known as transceiver signal linearization, which can
facilitate decoding. This linearization allows the code to be
backward compatible with existing signal processing techniques and
standards, and allows for the design of low complexity interference
suppressing filters and channel equalizers \cite{SBP04}. It has been
shown that a complex orthogonal design can achieve transceiver
signal linearization if and only if each row in the code has either
all conjugated entries or all non-conjugated entries \cite{SBP04},
which is called conjugation separated. Note that $\mathcal{G}_n$ and
$\mathcal{G}^w_n$ are all conjugation separated and thus satisfy the
transceiver signal linearization property.

When $n\equiv 0 \pmod 4$, it's possible to pad an extra column on
$\mathcal{G}_{n-1}$ to obtain a new COD.

\begin{theorem}
\label{thm:On} For positive integer $n=2m, m$ even, let
$\mathcal{H}_n = (\mathcal{G}_{n-1}, \mathcal{L}_n),$ where
$\mathcal{L}_n(\alpha) = \alpha(n)(-1)^{\psi(\alpha)}z_{\alpha\oplus
e_n}, \text{ for all }\alpha \in \mathbb{F}_2^n$ and
$\psi(\alpha)=\sum_{i=1}^m{\alpha(2i)}$. Then $\mathcal{H}_n$ is a
COD with parameter $[2^n, n, 2^{n-1}]$.
\end{theorem}
\begin{proof} From Theorem \ref{thm:Gn}, we claim columns in
$\mathcal{G}_{n-1}$ are pairwise orthogonal. By proving
$\mathcal{L}_{n}$ is orthogonal to the other columns, we can
complete the proof. It's obvious that each variable exists on
$\mathcal{H}_n$ only once. It only remains to prove any two nonzero
elements (one is on $\mathcal{L}_n$) in the same row are in an
Alamouti $2 \times 2$.

Consider column $1 \leq i \leq n - 1$ and column $n$. For row
$\alpha\in\mathbb{F}_2^n$ and $\alpha(i)=\alpha(n) = 1$,
$\mathcal{H}_{n}(\alpha, i) = (-1)^{\theta(\alpha,
i)}z^*_{\varphi(\alpha,i)}$ and $\mathcal{H}_{n}(\alpha, n) =
(-1)^{\psi(\alpha)}z_{\alpha \oplus e_n}$. Since each variable
exists in $\mathcal{L}_{n}$, there exists an integer $\beta \in
\mathbb{F}^n_2, \beta(n)=1$ such that $|\mathcal{H}_{n}(\beta,
n)|=|\mathcal{H}_{n}(\alpha, i)|^*$, which is
\begin{equation}
\beta \oplus e_n = \alpha \oplus \alpha(n)e \oplus e_i. \nonumber
\end{equation}
Noting that $\alpha(n)=1$, thus,
\begin{equation}
\label{equ:sum_s_s'_2} \beta = \alpha \oplus e \oplus e_i \oplus
e_n.
\end{equation}
which implies $\beta(i) = \beta(n) = 1$. Now, we calculate the
subscript of the variable in $\mathcal{H}_{n}(\beta, i)$.
\begin{eqnarray}
\varphi(\beta, i) & = & \beta \oplus \beta(n)e \oplus e_i \nonumber \\
& = & \alpha \oplus e \oplus e_i \oplus e_n \oplus e \oplus e_i
\nonumber \\
& = & \alpha \oplus e_n, \nonumber
\end{eqnarray}
which is equal to the subscript of variable in
$\mathcal{H}_{n}(\alpha, n)$. Therefore, the submatrix
$\mathcal{H}_{n}(\alpha, \beta; i, n)$ could be written as follows
\begin{displaymath}
\begin{pmatrix}
(-1)^{\theta(\alpha, i)} z^*_{\gamma} & (-1)^{\psi(\alpha)}z_{\delta} \\
(-1)^{\theta(\beta, i)} z^*_{\delta} & (-1)^{\psi(\beta)}z_{\gamma},
\end{pmatrix},
\end{displaymath}
where $\gamma = \alpha \oplus e_i \oplus e_n$ and $\delta = \alpha
\oplus e_n$. Let's check the signs to verify whether it's an
Alamouti $2 \times 2$.

When $i$ is even,
\begin{align*}
& \:\: \theta(\alpha,i)+\psi(\alpha)+\theta(\beta,i)+\psi(\beta) \\
\equiv & \: \wt_{i, n}(\alpha) + \frac{i}{2} +
\sum_{k = 0}^{m}{\alpha(2k)} + \\
 & \: \wt_{i, n}(\beta) + \frac{i}{2} + \sum_{k = 1}^{m}{\beta(2k)} && \text{by definition} \\
\equiv & \: \wt_{i, n}(\alpha \oplus \beta) + \sum_{k = 1}^{m}{(\alpha(2k) \oplus \beta(2k))} &&\\
\equiv & \: (n - i-1) + (m - 1) && \text{by \eqref{equ:sum_s_s'_2}} \\
\equiv & \: 1 \pmod{2}.
\end{align*}

When $i$ is odd,
\begin{align*}
& \:\: \theta(\alpha,i)+\psi(\alpha)+\theta(\beta,i)+\psi(\beta) \\
\equiv & \: \wt_{i, n}(\alpha) + \frac{i-1}{2} + \alpha(n) +
\sum_{k = 1}^{m}{\alpha(2k)} + \\
& \: \wt_{i, n}(\beta) + \frac{i-1}{2} + \beta(n) + \sum_{k = 1}^{m}{\beta(2k)} && \text{by definition}\\
\equiv & \: \wt_{i, n}(\alpha \oplus \beta) + (\alpha\oplus\beta)(n) + \sum_{k = 1}^{m}{(\alpha(2k) \oplus \beta(2k))} &&\\
\equiv & \: (n - i - 1) + 0 + (m - 2) && \text{by \eqref{equ:sum_s_s'_2}} \\
\equiv & \: 1 \pmod{2}.
\end{align*}
Therefore,
\begin{displaymath}
(-1)^{\theta(\alpha,i)}z_\gamma(-1)^{\psi(\alpha)}z_\delta +
(-1)^{\theta(\beta, i)}z_\delta(-1)^{\psi(\beta)}z_\gamma = 0
\end{displaymath}
holds and the submatrix $\mathcal{H}_{n}(\alpha, \beta; i, n)$ an
Alamouti $2 \times 2$, which implies column $i$ and column $n$ are
orthogonal.
\end{proof}

Similar with the idea in Theorem \ref{thm:Gnw_main}, by taking out
some submatrices of $\mathcal{H}_n$, we can obtain new ones.

\begin{theorem}
\label{thm:Onm} For positive integer $n=2m, m$ even, let $$
\mathcal{H}^m_n = \mathcal{H}_n(\alpha_1, \ldots,
\alpha_{\binom{n}{m+1}}; 1,\ldots,n),
$$ where $\alpha_i$ are all vectors in $\mathbb{F}^n_2$ with weight
$m+1$. Then $\mathcal{H}^m_n$ is a COD with parameter $[{n \choose
m+1}, n, {n-1 \choose m}]$.
\end{theorem}
\begin{proof}
From Theorem \ref{thm:On}, we know $\mathcal{H}_{n}$ is orthogonal.
Now we will prove that every variable with subscript weight $m$
exists on each column, which implies $\mathcal{H}^m_{n}$ is a COD.

For $\alpha \in \mathbb{F}^n_2, \alpha(n)=0, 1 \leq i \leq n-1$ and
$\alpha(i) = 1$, since $\wt(\varphi(\alpha, i)) = \wt(\alpha \oplus
e_i) = \wt(\alpha) - 1$, then $\wt(\varphi(\alpha, i)) = m$ if and
only if $\wt(\alpha) = m+1$.

For $\alpha \in \mathbb{F}^n_2, \alpha(n)=1, 1 \leq i \leq n-1$ and
$\alpha(i) = 1$, since $\wt(\varphi(\alpha, i)) = \wt(\alpha \oplus
e_i \oplus e) = n+1 - \wt(\alpha)$, then $\wt(\varphi(\alpha, i)) =
m$ if and only if $\wt(\alpha) = n+1-m=m+1$.

For the last column, since $\mathcal{L}_n(\alpha) =
(-1)^{\psi(\alpha)}z_{\alpha\oplus e_n}$ for $\alpha(n)=1$, it's
easy to see if $\wt(\alpha ) = m+1$, then $\wt(\psi(\alpha)) = 2m$
and vice versa.
\end{proof}

It's worth noticing that, for a given row, there are both conjugated
and non-conjugated nonzero entries in $\mathcal{H}_n$ and
$\mathcal{H}^m_n$, which violets the transceiver signal
linearization property.

In \cite{AKM11}, Adams et al. proved that when $n \equiv 1,2,3 \pmod
4$, maximal rate CODs with transceiver linearization can achieve the
minimal delay, and when $n \equiv 0 \pmod 4$, it can not. Our
explicit-form constrictions are consistent with their results.

The CODs constructed by Liang in \cite{Lia03}, and by Su and Xia in \cite{SX02} is exactly $\mathcal{G}_n^m$, which achieves
maximal rate and minimal delay when $n \not \equiv 0 \pmod 4$. The closed-form constructions
in \cite{LFX05} are exactly $\mathcal{G}_n^m$ and $\mathcal{H}^m_n$, and therefore achieve maximal rate
and minimal delay for any $n$. The constructions in \cite{ADK11} by Adams et al. have rate $1/2$ and delay $2^{m-1}$ or $2^m$, depending on
the parity of $n$ modulo $8$. Those CODs do not belong to first type, and have smaller decoding delay compared to $\mathcal{G}^w_n$ with rate near $1/2$.

\section{Structures of Atomic first type CODs}
In \cite{AKM10}, it is proved that in a COD with parameter
$[{2m\choose m-1},n,{2m-1 \choose m-1}]$ when $n=2m$ or $2m-1$, row
$\alpha$ and row $\beta$ share an Alamout $2 \times 2$ over column
$i$ and $j$ if and only if the zero pattern of row $\alpha$ and row
$\beta$ are simultaneously nonzero exactly in columns $i$ and $j$
and never simultaneously zero or nonzero in any other column. In
fact, it can be generalized for first type COD as follows.

\begin{lemma}
\label{lem:0pat_to_sc} For first type COD $\mathcal{O}_z[p, n, k]$,
$\mathcal{O}_z(\alpha, i)$ and $\mathcal{O}_z(\beta, j)$ are the
same, up to signs, implies that the zero pattern of row $\alpha$ and
that of row $\beta$ are different only at column $i$ and column $j$;
$\mathcal{O}_z(\alpha, i)$ and $\mathcal{O}_z(\beta, j)$ are
conjugated, up to signs, implies that the zero pattern of row
$\alpha$ and that of row $\beta$ are the same only at column $i$ and
column $j$.
\end{lemma}
\begin{proof}
Without loss of generality, assume $\mathcal{O}_z(\alpha, i)=z[1]$
and $\mathcal{O}_z(\beta, j)=z[1]$, where $z[1]$ represents an
arbitrary element in $\{z_1,-z_1,z^*_1,-z^*_1\}$. Through some
column permutation, say $\pi \in \Sigma_n$, where $\Sigma_n$ is the
set of all permutations on $\{1,2,\ldots,n\}$, we can transform
$\mathcal{O}_z$ into $\mathcal{B}_1$ form, where
\begin{eqnarray}
\mathcal{B}_1 & = & \begin{pmatrix} z_1 I_{n_1} & \mathcal{M}_1 \\
-\mathcal{M}^H_1 & z^*_1 I_{n-n_1}
\end{pmatrix} \nonumber\\
& = &
       \left(
       \begin{array}{c|c}
       \begin{array}{cccc}
       z_1&0&\cdots&0\\
       0&z_1&\cdots&0\\
       \vdots&\vdots&\ddots&\vdots\\
       0&0&\cdots&z_1
       \end{array}&\mathcal{M}_1\\
       \hline
       -\mathcal{M}_1^H&\begin{array}{cccc}
       z_1^*&0&\cdots&0\\
       0&z_1^*&\cdots&0\\
       \vdots&\vdots&\ddots&\vdots\\
       0&0&\cdots&z_1^*
       \end{array}
       \end{array}
       \right), \label{equ:Bj_form}
\end{eqnarray}
and $\mathcal{M}_1$ contains no zero entry.

When $|\mathcal{O}_z(\alpha, i)|=|\mathcal{O}_z(\beta, j)|$, we know
row $\alpha$ and row $\beta$ are both in the upper or lower part of
$\mathcal{B}_1$ form. We can see that row $\alpha$ and row $\beta$
have the same zero pattern except for column $\pi(i)$ and $\pi(j)$
after column permutation, which implies that row $\alpha$ and row
$\beta$ have the same zero pattern except for column
$\pi^{-1}(\pi(i))=i$ and $\pi^{-1}(\pi(j))=j$ before column
permutation.

When $|\mathcal{O}_z(\alpha, i)|=|\mathcal{O}_z(\beta, j)|^*$, we
know row $\alpha$ and row $\beta$ are in different parts (upper or
lower) of $\mathcal{B}_1$ form. We can see that the zero patterns of
row $\alpha$ and row $\beta$ are all different except for column
$\pi(i)$ and $\pi(j)$ after column permutation, which implies the
zero patterns of row $\alpha$ and row $\beta$ are all different
except for column $\pi^{-1}(\pi(i))=i$ and $\pi^{-1}(\pi(j))=j$
before column permutation.
\end{proof}

The next lemma states that in a first type COD, the existence of one
zero pattern implies the existence of some other zero patterns,
which will be used to prove the lower bound of decoding delay $p$
for first type COD.

\begin{lemma}
\label{lem:01patern_transp} Let $\mathcal{O}_z[p,n,k]$ be a first type
COD. If one zero pattern of some row is $\alpha \in \mathbb{F}_2^n$,
then for any $1\le i \not= j \le n$, there exists one row with zero
pattern $\beta \in \mathbb{F}_2^n$, such that $\beta(i) = \alpha(j),
\beta(j) = \alpha(i)$ and $\beta(l) = \alpha(l)$ for all $l \not= i,
j$.

Furthermore, for any $1 \le i \not= j \le n$ such that $\alpha(i) =
\alpha(j) = 1$, there exists one row with zero pattern $\beta \in
\mathbb{F}_2^n$, such that $\beta = \alpha \oplus e_i \oplus e_j
\oplus e$.
\end{lemma}
\begin{proof}
\textbf{For the first part:} as when $\alpha(i)=\alpha(j)$, the
conclusion is trivial, we assume $\alpha(i)=1$ and $\alpha(j)=0$.
And, without loss of generality, assume the variable on that row in
column $i$ is $z[1]$. Through column permutation $\pi$ satisfying
$\pi(i)=1, \pi(j)=2$ and
$\pi(\alpha)=(1,\underbrace{0,\ldots,0}_{n-\wt(\alpha)},1,\ldots,1)$,
we can make this row the first row in $\mathcal{B}_1$ form.

Recall $\mathcal{B}_1$ form \eqref{equ:Bj_form}, we know the zero
pattern of the second row is different from $\pi(\alpha)$ only in
column 1 and 2, which implies that it's different from $\alpha$ only
in column $\pi^{-1}(1)=i$ and $\pi^{-1}(2)=j$ before column
permutation.

\textbf{For the second part:} Without loss of generality, assume the
variable on that row in column $i$ is $z[1]$. Through column
permutation $\pi$ such that $\pi(i)=1, \pi(j)=n-\wt(\alpha)+2$ and
$\pi(\alpha)=(1,\underbrace{0,\ldots,0}_{n-\wt(\alpha)},1,\ldots,1)$,
we can make this row the first row in $\mathcal{B}_1$ form after
column permutation. Recall $\mathcal{B}_1$ form \eqref{equ:Bj_form},
we know the zero pattern of the first row of the lower part is the
same as $\pi(\alpha)$ only in column 1 and $n-\wt(\alpha)+2$, which
implies that it's only the same as $\alpha$ in column
$\pi^{-1}(1)=i$ and $\pi^{-1}(n-\wt(\alpha)+2)=j$ before column
permutation.
\end{proof}

Next lemma gives an lower bound of the decoding delay $p$ for first type
COD when $n$ and the number of nonzero entries in some row are
given.

\begin{lemma}
\label{lem:lowbound_p_nw} Let $\mathcal{O}_z[p,n,k]$ be a first type
COD. If one row in $\mathcal{O}_z$ contains $w+1$ nonzero entries,
then $ p \ge {n \choose w-1} + {n \choose w+1}$ when $n \not= 2w$;
and $p \ge {n \choose w-1}$ when $n = 2w$.

Furthermore, all zero patterns with weight $w+1$ or $n-w+1$ exists
in $\mathcal{O}_z$.
\end{lemma}
\begin{proof}
According to the condition, assume that one row in $\mathcal{O}_z$
has zero pattern $\alpha \in \mathbb{F}_2^n$ such that
$\wt(\alpha)=w+1$. Then for any zero pattern $\beta \in
\mathbb{F}_2^n$ with $\wt(\beta)=w+1$, there exists a permutation
$\pi \in \Sigma_n$ such that $\pi(\alpha) = \beta$. Since any
permutation is a product of transpositions, then $\pi$ can be
written as the product of transpositions. According to Lemma
\ref{lem:01patern_transp}, we claim there exists one row in
$\mathcal{O}_z$ with zero pattern $\beta = \pi(\alpha)$.

Again, by Lemma \ref{lem:01patern_transp}, the existence of zero
pattern $\alpha$ implies one row with zero pattern $\beta$ such that
$\wt(\beta) = n+2-\wt(\alpha)=n-w+1$. By similar arguments in the
last paragraph, we claim all zero patterns with weight $n-w+1$
exist. When $w+1 \not= n-w+1 \Leftrightarrow n \not= 2w$, we know
$p$ is lower bounded by the number of all zero patterns with weight
$w+1$ and $n-w+1$, i.e., $p \ge \binom{n}{w+1} + \binom{n}{n-w+1}$.
When $w+1 = n-w+1 \Leftrightarrow n = 2w$, we know $p$ is lower
bounded by the number of all zero patterns with weight $w+1$, i.e.,
$p \ge \binom{n}{w+1}=\binom{n}{w-1}$.
\end{proof}

For first type COD, besides the lower bound, we can say more about the
decoding delay $p$, as the following lemma reveals.

\begin{lemma}
 Let $\mathcal{O}_z[p,n,k]$ be an atomic first type COD. If
one row in $\mathcal{O}_z$ contains $w+1$ nonzero entries, then $p$
is a multiple of ${n \choose w-1} + {n \choose w+1}$ when $n \not=
2w$; and $p$ is a multiple of ${n \choose w-1}$ when $n = 2w$.
\end{lemma}

\begin{proof}
At first, we will show, for an atomic first type COD
$\mathcal{O}_z[p,n,k]$, if one row contains $w+1$ nonzero entries,
then each row contains $w+1$ or $n-w+1$ nonzero entries. Since
$\mathcal{O}_z$ is atomic, then, for any pair of $1 \le s \not= t
\le k$, there exists $j_1=s, j_2, \ldots, j_m=t$ such that
$\mathcal{B}_{j_1}$ and $\mathcal{B}_{j_2}$ share some common rows,
$\mathcal{B}_{j_2}$ and $\mathcal{B}_{j_3}$ share some common rows,
..., $\mathcal{B}_{j_{m-1}}$ and $\mathcal{B}_{j_m}$ share some
common rows. Note that, in some $\mathcal{B}_j$ form of first type COD,
if one row contains $w+1$ nonzero entries, then all rows in
$\mathcal{B}_j$ contains $w+1$ or $n-w+1$ nonzero entries. As $s$
and $t$ are taken arbitrarily, we claim every row of
$\mathcal{O}_z[p,n,k]$ contains $w+1$ or $n-w+1$ nonzero entries.

Assume that zero pattern $\alpha$ appears with maximal times $t$,
say, row $r_1, \ldots, r_t$ have zero pattern $\alpha$. For any $i$
satisfying $\alpha(i)=1$, there exists a column permutation $\pi$ on
$\mathcal{O}_z$ such that $\pi(i)=1$ and $\pi(\alpha)=(1,
\underbrace{0, \ldots, 0}_{n-w-1}, 1, \ldots, 1)$. Therefore, $r_i$
is in $\mathcal{B}_{j_i}$ form, where $z[j_i]$ appears in the first
column of row $r_i$ after column permutation $\pi$. Since $z[j_i]$
appears in the same column, $z[j_1], \ldots, z[j_t]$ are all
different, and thus form $\mathcal{B}_{j_i}$ are mutually
disjointed.

Now, we will show all zero patterns with weight $w+1$ exist $t$
times. Recall $\mathcal{B}_j$ form \eqref{equ:Bj_form}, we claim
there are $t$ different rows with zero pattern $\beta$, where
$\beta$ is obtained by exchanging the value on $i^{\text{th}}$ and
$j^{\text{th}}$ of $\alpha$ with any $\alpha(i)\oplus\alpha(j)=1$.
Since any permutation can be written as the product of
transpositions, repeat this procedure, we know all zero patterns
with weight $w+1$ exists at least $t$ times. By the maximality of
$t$, we claim all zero patterns with weight $w+1$ exists $t$ times.

Finally, we will show all zero patterns with weight $n-w+1$ exist
$t$ times. For any $j, \alpha(j)=1$, recall $\mathcal{B}_j$ form
\eqref{equ:Bj_form}, we claim there are $t$ different rows with zero
pattern $\beta=\alpha \oplus e_i \oplus e_j \oplus e$. Following
similar argument of the above paragraph, we claim all zero patterns
with weight $n-w+1$ exists $t$ times.

Therefore, we have $p = t\left({n \choose w-1} + {n \choose
w+1}\right)$ when $n \not= 2w$; and $p = t {n \choose w-1}$ when $n
= 2w$, where $t$ is a positive integer.
\end{proof}

Next three lemmas are about the structure of COD $\mathcal{G}^w_n$
and $\mathcal{H}^m_n$, and they will be used in the proof of Theorem
\ref{thm:Gnw_main}.

\begin{lemma}
\label{lem:smalst_idx_row} For $\alpha \in \mathbb{F}_2^{n+1},
\alpha(n+1)=0$ and $1 \le i \le n$, $z[\alpha]$ is the variable with
smallest index on row $\alpha_i$ of $\mathcal{G}_n$, where
$\alpha_i=\alpha \oplus e_i \oplus \alpha(i)e$, if and only if
\begin{equation}
\alpha(i)=\alpha(i+1)=\ldots=\alpha(n)=0,
\end{equation}
or
\begin{equation}
\alpha(1)=\alpha(2)=\ldots=\alpha(i)=1.
\end{equation}
\end{lemma}
\begin{proof}
By the definition of $\mathcal{G}^w_n$, we know that for all $j$
satisfying $\alpha_i(j)=1 \Leftrightarrow \alpha(i)\oplus
\alpha(j)=1$, $\mathcal{G}^w_n(\alpha_i, j)=z[\beta]$, where
$\beta=\alpha_i \oplus e_j \oplus \alpha_i(n+1)e=\alpha \oplus e_i
\oplus \alpha(i)e \oplus e_j \oplus \alpha(i)e =\alpha\oplus e_i
\oplus e_j$.

Therefore, if $\alpha(i)=0$, for any $j, \alpha(j)=1$, $\alpha\oplus
e_i \oplus e_j > \alpha$ if and only $j < i \Rightarrow$
$\alpha(i)=\alpha(i+1)=\ldots=\alpha(n)=0$; if $\alpha(i)=1$, for
any $\alpha(j)=0$, $\alpha\oplus e_i \oplus e_j > \alpha$ if and
only if $j > i \Rightarrow\alpha(1)=\alpha(2)=\ldots=\alpha(i)=1$.

\end{proof}

\begin{lemma}
\label{lem:smalst_idx_Bj}
 For $\alpha \in \mathbb{F}_2^{n+1},
\alpha(1)=\ldots=\alpha(s-1)=1, \alpha(s)=0, \alpha(t)=1,
\alpha(t+1)=\ldots=\alpha(n+1)=0$ and $1\le s<t\le n$, the smallest
index of variables in $\mathcal{B}_{\alpha}$ form of $\mathcal{G}_n$
is $\alpha\oplus e_s \oplus e_t$.
\end{lemma}
\begin{proof} We prove it by calculating the indexes of all variables
in $\mathcal{B}_{\alpha}$ form directly. By the definition of
$\mathcal{G}_n$, we know $z[\alpha]$ is in $\mathcal{G}_n(\alpha_i,
i), i=1,2,\ldots,n$, where $\alpha_i = \alpha \oplus e_i \oplus
\alpha(i)e$. For $\alpha_i(j)=1 \Leftrightarrow \alpha(i) \oplus
\alpha(j)=1$, $\mathcal{G}_n(\alpha_i, j)=z[\beta]$, where
$\beta=\alpha_i \oplus e_j \oplus \alpha_i(n+1)e = \alpha \oplus e_i
\oplus \alpha(i)e \oplus e_j \oplus \alpha(i)e=\alpha \oplus e_i
\oplus e_j$.

For what $i, j$ satisfying $\alpha(i)\oplus\alpha(j)=1$, value
$\alpha \oplus e_i \oplus e_j$ reaches the minimal? Without loss of
generality, assume $\alpha(i)=0$ and $\alpha(j)=1$. It's easy to see
that $i$ should be as small as possible and $j$ should be as big as
possible. Therefore, $i=s$ and $j=t$, and $\alpha\oplus e_s \oplus
e_t$ is the smallest index of variables in $\mathcal{B}_{\alpha}$
form of $\mathcal{G}_n$.
\end{proof}

\begin{lemma} For $n=2m$ or $2m-1$, CODs $\mathcal{G}^w_n$, $-1 \le w \le n+1$, are all
atomic. And when $n=2m$, $m$ even, COD $\mathcal{H}^m_n$ is atomic.
\end{lemma}
\begin{proof}
If $n \not= 2w$, $\mathcal{G}^w_n$ has parameter $[{n \choose
w-1}+{n \choose w+1}, n, {n \choose w}]$. By Lemma
\ref{lem:lowbound_p_nw}, we know $p$ is minimal, and therefore
$\mathcal{G}^w_n$ is atomic. It's similar to prove $\mathcal{H}^m_n$
is atomic, for $n=2m$, $m$ even.

For $n=2m$, COD $\mathcal{G}^m_n$, assume that there is an atomic
COD $\mathcal{O}_z$ consisting of some rows of $\mathcal{G}^m_n$.
Take one $\alpha \in \mathbb{F}_2^{n+1}, \wt(\alpha)=m+1$ and
$\alpha(n+1)=0$, such that $z[\alpha]$ appears in $\mathcal{O}_z$.
By the definition of $\mathcal{G}^m_n$, $z[\alpha]$ appears in
$\mathcal{G}^m_n(\alpha_i, i)$, where $\alpha_i = \alpha \oplus e_i
\oplus \alpha(i)e$. For any $j$, satisfying $\alpha_i(j)=1
\Leftrightarrow \alpha(i) \oplus \alpha(j)=1$,
$\mathcal{G}^m_n(\alpha_i, j)$ contains the variable with index
$\alpha_i \oplus e_j \oplus \alpha_i(n+1)e = \alpha \oplus e_i
\oplus \alpha(i)e \oplus e_j \oplus \alpha(i)e = \alpha \oplus e_i
\oplus e_j$. Thus $z[\alpha \oplus e_i \oplus e_j]$ should exist in
$\mathcal{O}_z$. Since $i$ and $j$ are taken arbitrary if
$\alpha(i)\oplus\alpha(j)=1$ is satisfied, by repeating this
procedure, we claim all variables with index weight
$\wt(\alpha)=m+1$ appears in $\mathcal{O}_z$. Therefore
$\mathcal{O}_z = \mathcal{G}^m_n$, and the proof is complete.
\end{proof}

The next theorem is our main result, which determines the parameters
as well as the structures of most atomic first type CODs.

\begin{theorem}
\label{thm:Gnw_main} Let $\mathcal{O}_z[p,n,k]$ be an atomic COD,
with some row containing $w+1$ nonzero entries, and $n \not= 2w$.
Then, $[p, n, k]=[{n \choose w-1} + {n \choose w+1}, n, {n \choose
w}]$ and $\mathcal{O}_z$ is the same as $\mathcal{G}^w_n$ under
equivalence operation.
\end{theorem}
\begin{proof}
We first present an example to illustrate our proof idea. For some
atomic COD $\mathcal{O}_z$, with $n = 3$ and $w = 2$, we will show
how to prove it is the same as
$$\mathcal{G}^2_3=
\begin{pmatrix}
-z_{(0,1,1)} & z_{(1,0,1)} & z_{(1,1,0)}\\
-z^*_{(1,0,1)} & -z^*_{(0,1,1)} & 0\\
-z^*_{(1,1,0)} & 0 & -z^*_{(0,1,1)}\\
0 & z^*_{(1,1,0)} & -z^*_{(1,0,1)}
\end{pmatrix}.
$$
For convenience, we denote $z_{(1,1,0)}$ by $z_1$, $z_{(1,0,1)}$ by
$z_2$, $z_{(0,1,1)}$ by $z_3$.

Since $w = 2$, there is at least one row of $\mathcal{O}_z$ which
contains $3$ nonzero entries. Without loss of generality, we denote
it by
$$
\begin{pmatrix}
- z_3 & z_2 & z_1
\end{pmatrix}.
$$
It can be achieved by instance renaming, instance conjugation and
instance negation.

Recalling $\mathcal{B}_1$ form, we claim there exists one row of
$\mathcal{O}_z$ matches $(\pm z^*_1, 0, \bigstar)$. At first, we can
use row negation to make sure $\pm z^*_1$ takes the same sign as
that of $\mathcal{G}^w_n$, which is ``$-$''. By orthogonality,
 the first row shares an Alamouti $2 \times 2$ with this row, which is $\begin{pmatrix} - z_3 &  z_1 \\
-z^*_1 & \bigstar \end{pmatrix}$. Thus, $\bigstar$ should be
$-z^*_3$. Now we have determined two rows of $\mathcal{O}_z$ as
follows
$$
\begin{pmatrix}
- z_3 &  z_2 & z_1 \\
- z^*_1 & 0 & -z^*_3
\end{pmatrix}.
$$

Recalling $\mathcal{B}_1$ form again, there must exist one row of
$\mathcal{O}_z$ matches $(0, \pm z^*_1, \bigstar)$. At first, we can
use row negation to make sure $\pm z^*_1$ takes the same sign as
that of $\mathcal{G}^w_n$, which is
``$+$''. By orthogonality, the first row shares an Alamouti $2 \times 2$ with this row, which is $\begin{pmatrix} z_2 &  z_1 \\
z^*_1 & \bigstar \end{pmatrix}$. Thus, $\bigstar$ should be $-
z^*_2$. Now, we have determined three rows of $\mathcal{O}_z$ as
follows
$$
\begin{pmatrix}
- z_3 &  z_2 &  z_1 \\
- z^*_1 & 0 & - z^*_3 \\
0 & z^*_1 & - z^*_2
\end{pmatrix}.
$$

Recalling $\mathcal{B}_2$ form, there must exist one row of
$\mathcal{O}_z$ matches $(\pm z^*_2, \bigstar, 0)$. At first, we can
use row negation to make sure $\pm z^*_1$ takes the same sign as
that of $\mathcal{G}^w_n$, which is
``$-$''. By orthogonality, the first row shares an Alamouti $2 \times 2$ with this row, which is $\begin{pmatrix} - z_3 &  z_2 \\
- z^*_2 & \bigstar \end{pmatrix}$, which implies $\bigstar$ should
be $- z^*_3$. Now, we have
$$
\begin{pmatrix}
- z_3 &  z_2 &  z_1 \\
- z^*_1 & 0 & - z^*_3 \\
0 & z^*_1 & - z^*_2 \\
- z^*_2 & - z^*_3 & 0
\end{pmatrix},
$$
which is already a COD. Since $\mathcal{O}_z$ is atomic, we claim
$[p, n, k] = [4, 3, 3]$ and it is the same as $\mathcal{G}^2_3$
under equivalence operation.

\vspace{0.3cm} Applying the above method, for a general
$\mathcal{O}_z[p, n, k]$ with some row containing $w+1$ nonzero
entries, $n \not= 2w$, we will prove that, using equivalence
operation, we can transform $\mathcal{O}_z$ to $\mathcal{G}^w_n$ row
by row in a specific order.

%Define $\alpha < \beta$ for $\alpha, \beta \in \mathcal{F}_2^n$ if
%and only if $\sum_{i=1}^{n}{\alpha(i) 2^i} < \sum_{i=1}^{n}{\beta(i)
%2^i}$ .

We reorder the rows in $\mathcal{G}^w_n$ first by order of the
smallest index of the variables on that row in increasing, then by
the order of the row index in increasing. We will use induction to
prove that,  $\mathcal{O}_z$ is the same as $\mathcal{G}^w_n$ under
equivalence operation, and the induction parameter is the reordered
rows of $\mathcal{G}^w_n$.

\textbf{Induction basis:} For the first row of $\mathcal{G}^w_n$,
say row $\beta \in \mathbb{F}_2^{n+1}$. In $\mathcal{G}^w_n$, find
one row with the zero pattern $(\beta(1), \beta(2), \ldots,
\beta(n))$. Note that Lemma \ref{lem:lowbound_p_nw} guarantees the
existence of this row. Since all variables exist for the first time,
we can use instance renaming, instance conjugation and instance
negation to make this row the same as the corresponding row of
$\mathcal{G}^w_n$.

\textbf{Induction step:} For variable index $\beta \in
\mathbb{F}_2^{n+1}, \beta(n+1)=0$ and row $\beta_i \Rightarrow
\mathcal{G}^w_n(\beta_i, i)=z[\beta]$, where $\beta_i = \beta \oplus
e_i \oplus \beta(i)e$. Assume that there exists an equivalence
operation on $\mathcal{O}_z$ such that some rows of $\mathcal{O}_z$
are the same as rows of $\mathcal{G}^w_n$ which either has the
smallest index less than $\beta$ or the smallest index $\beta$ and
its row index less than $\beta$, we will show it is true after row
$\beta_i$ in $\mathcal{G}^w_n$ is added.

We claim $z[\beta]$ already exists in former induction steps.
Otherwise, $\beta$ should have the smallest index of all variables,
and Lemma \ref{lem:smalst_idx_Bj} implies $\beta$ is unique and thus
already appears on the first row in the induction step. Therefore,
by Lemma \ref{lem:lowbound_p_nw}, we know that there exists one row
of $\mathcal{O}_z$ having the same zero pattern as row $\beta_i$ of
$\mathcal{G}^w_n$ with the corresponding position occupied
$z[\beta]$. Since $z[\beta]$ already exists, whether $z[\beta]$
takes conjugation is already determined by the zero pattern
$(\beta_i(1), \beta_i(2), \ldots, \beta_i(n))$. Thus, we can use row
negation to make sure $z[\beta]$ takes the same sign the same as
$\mathcal{G}^w_n(\beta_i, i)$. We will show that for all the other
nonzero entries on this row of $\mathcal{O}_z$,
\begin{itemize}
\item
either the variable exists for the first time (and it can't be a used variable), which implies we can
use instance renaming, instance conjugation and instance negation to
make it the same as the corresponding one in $\mathcal{G}^w_n$,
\item
or it's  uniquely determined, including sign and conjugation, by the
orthogonality of $\mathcal{O}_z$.
\end{itemize}

For any $j \not= i, \beta_i(j)=1 \Leftrightarrow
\beta(i)\oplus\beta(j)=1$, let's consider the entry on
$\beta_i^{\text{th}}$ row and $j^{\text{th}}$ column of
$\mathcal{O}_z$. By assumption that $z[\beta]$ is the smallest-index
variable in row $\beta_i$, we know from Lemma
\ref{lem:smalst_idx_row}, either
\begin{equation}
\label{equ:beta_eq0} \beta(i)=\beta(i+1)=\ldots=\beta(n)=0,
\end{equation}
or
\begin{equation}
\label{equ:beta_eq1} \beta(1)=\beta(2)=\ldots=\beta(i)=1
\end{equation}
holds. We will discuss it in the following four cases separately.
%\begin{itemize}
%\item $\beta(i)=\beta(i+1)=\ldots=\beta(n)=0$ and $\beta(l)=0$ for
%some $1\le l < j$.
%\item $\beta(i)=\beta(i+1)=\ldots=\beta(n)=0$ and
%$\beta(1)=\beta(2)=\ldots=\beta(j)=1$.
%\item $\beta(1)=\beta(2)=\ldots=\beta(i)=1$ and
%$\beta(l)=1$ for some $1\le l > j$
%\item
%$\beta(1)=\beta(2)=\ldots=\beta(i)=1$ and
%$\beta(j)=\beta(j+1)=\ldots=\beta(n)=0$.
%\end{itemize}

\textbf{Case 1: $\beta(i)=\beta(i+1)=\ldots=\beta(n)=0, \beta(j)=1$
and $\beta(l)=0$ for some $1\le l < j$.} Let
$\mathcal{G}^w_n(\beta_j, l)=z[\gamma]$, where $\beta_j=\beta\oplus
e_j\oplus e$. We have
\begin{eqnarray}
\gamma & = & \beta_j \oplus e_l \oplus \beta_j(n+1)e \nonumber\\
& = & \beta\oplus e_j\oplus e \oplus e_l \oplus e \nonumber\\
& = & \beta\oplus e_j \oplus e_l. \nonumber
\end{eqnarray}
Since $\beta(l)=0, \beta(j)=1$ and $l < j$, we have $\gamma <
\beta$, which implies row $\beta_j$ is already determined. By the
orthogonality of $\mathcal{O}_z$, submatrix
$$\mathcal{O}_z(\beta_i, \beta_j;i, j) =
\begin{pmatrix}
z[\beta] & \bigstar \\
z[\gamma] & z[\beta]
\end{pmatrix}
$$
should be an Alamouti $2 \times 2$. Thus $\bigstar$ should be
$z[\gamma]$ and its conjugation and sign are uniquely determined by
the other three entries.

 \textbf{Case 2:
$\beta(i)=\beta(i+1)=\ldots=\beta(n)=0$ and
$\beta(1)=\beta(2)=\ldots=\beta(j)=1$.} Let
$\mathcal{G}^w_n(\beta_i, j)=z[\gamma]$, where $\beta_i = \beta
\oplus e_i$. Thus $\gamma = \beta_i\oplus e_j \oplus \beta_i(n+1)
=\beta \oplus e_i\oplus e_j$.

To prove $z[\gamma]$ exists for the first time, it's sufficient to
show that there is no determined rows with zero pattern matching one
in $\mathcal{B}_{\gamma}$ form. Since $n \not= 2w$, any fixed row has a
unique zero pattern. By Lemma \ref{lem:smalst_idx_Bj}, we
know that $z[\beta]$ is the smallest index variable in $B_{\gamma}$
form of $\mathcal{G}^w_n$, adding the fact that $\beta_j =
\beta\oplus e_j \oplus e > \beta_i$, which implies that $z[\gamma]$
exists for the first time. Therefore, we can use instance renaming,
instance conjugation and instance negation to make
$\mathcal{O}_z(\alpha_i, j)$ the same as $\mathcal{G}^w_n(\alpha_i,
j)$.

\textbf{Case 3: $\beta(1)=\beta(2)=\ldots=\beta(i)=1, \beta(j)=0$
and $\beta(l)=1$ for some $l > j$.} Let
$\mathcal{G}^w_n(\beta_j, l)=z[\gamma]$, where $\beta_j=\beta\oplus
e_j$. We have $ \gamma  =  \beta_j \oplus e_l \oplus \beta_j(n+1)e =
 \beta\oplus e_j \oplus e_l. $ Since $\beta(l)=1, \beta(j)=0$ and
$l > j$, we have $\gamma < \beta$, which implies row $\beta_j$ is
already determined. Following the same argument in \textbf{Case 1},
we know $\mathcal{O}_z(\beta_j, j)$ is uniquely determined.

 \textbf{Case 4:
$\beta(1)=\beta(2)=\ldots=\beta(i)=1, \beta(j)=0$ and
$\beta(j)=\beta(j+1)=\ldots=\beta(n)=0$.} Let
$\mathcal{G}^w_n(\beta_j, j)=z[\gamma]$, where $\beta_j=\beta\oplus
e_j$. We have $ \gamma  =  \beta_j \oplus e_j \oplus \beta_j(n+1)e =
 \beta\oplus e_j \oplus e_j =\beta.$ Note that $\beta_j=\beta\oplus
e_j$ and $\beta_i = \beta\oplus e_i\oplus e$, which implies $\beta_j
< \beta_i$. Therefore, row $\beta_j$ in $\mathcal{O}_z$ is
determined. Following the same argument in \textbf{Case 1}, we know
element in $\mathcal{O}_z(\beta_i, j)$ is uniquely determined.

\end{proof}

%\begin{theorem}
%Let $\mathcal{O}_z[p,n,k]$ be an atomic COD, with some row
%containing $w+1$ nonzero entries, and $n \not= 2w$. Then
%$\mathcal{O}_z$ is the same as $\mathcal{G}^w_n$ under equivalence
%operation.
%\end{theorem}
%\begin{proof}
%In Theorem \ref{thm:Gnw_para}, we have proved that, up to signs,
%$\mathcal{O}_z$ is the same as $\mathcal{G}^w_n$ under equivalence
%operation. Now, we will show that, using row negation and instance
%negation, $\mathcal{O}_z$ can be turned into a unique form, which
%implies that $\mathcal{O}_z$ and $\mathcal{G}^w_n$ are the same
%under equivalence operation.

%Let's fix the signs of variables in the increasing order of variable
%index.

%For variable index $\beta \in \mathbb{F}_2^{n+1}, \beta(n+1)=0$,
%assume that all signs of $z[\beta']$ with $\beta' < \beta$ are
%fixed.

%For one row containing $z[\beta]$ with the smallest index, we can
%use row negation to make sure it is positive.

%Assume that $\mathcal{O}_z(\beta_i,i)=z[\beta]$, which implies
%$\beta=\beta_i \oplus e_i \oplus \beta_i(n+1)e$ and $\beta_i(i)=1$.
%Thus we have $\beta_i=\beta\oplus e_i \oplus \beta(i)e$. Just as we
%have indicated in Theorem \ref{thm:Gnw_para}, $z[\beta]$ is the
%variable with smallest index implies that either
%\begin{equation}
%\beta(i)=\beta(i+2)=\ldots=\beta(n)=0,
%\end{equation}
%or
%\begin{equation}
%\beta(1)=\beta(2)=\ldots=\beta(i)=1.
%\end{equation}
%Let $\beta(1)=\beta(2)=\ldots=\beta(s-1)=0, \beta(s)=1$ and $\beta($

%For those $z[\beta]$ which are not $z[\beta]$
%\end{proof}

It is worth noting that the equivalence operations used in transform
$\mathcal{O}_z$ to $\mathcal{G}^w_n$ does not contain column
negations. This property will be used in the sequel.

Theorem \ref{thm:Gnw_main} does not consider the case when $n=2w$.
To cover the final case, we need the following lemma first, which
states when $n\equiv2\pmod 4$, COD with parameter $[\binom{n}{m-1},
n, \binom{n-1}{m-1}]$ does not exist. This
result is first proved in \cite{AKM10}. However, based on our
explicit construction, we present another proof here.

\begin{lemma}
\label{lem:m_odd_notexist} When $n=2m$, $m$ odd, there does not
exist COD with parameter $[p, n, k] = [\binom{n}{m-1}, n,
\binom{n-1}{m-1}]$.
\end{lemma}
\begin{proof}
Assume to the contrary that there exists COD
$\mathcal{O}_z[\binom{n}{m-1}, n, \binom{n-1}{m-1}]$. By deleting
the last column of $\mathcal{O}_z$, we obtain a COD with parameter
$[\binom{n}{m-1}, n-1, \binom{n-1}{m-1}]$. By Theorem
\ref{thm:Gnw_main}, it is equivalent to $\mathcal{G}^m_{n-1}$. Thus,
by padding a column $\mathcal{L}$ to $\mathcal{G}^m_{n-1}$, we can
obtain a COD with parameter $[\binom{n}{m-1}, n, \binom{n-1}{m-1}]$.
Now, we will show it is impossible.

By Lemma \ref{lem:lowbound_p_nw}, we know the zero pattern of
$\mathcal{O}_z$ is unique. By Lemma \ref{lem:0pat_to_sc}, we know
which variable should be in $\mathcal{L}(\alpha)$ is uniquely
determined by the zero pattern of this row. Set
$\mathcal{L}(\alpha)=\alpha(n+1)\phi(\alpha)z_{\alpha \oplus
e_{n}}$, where $\phi(\alpha) \in \{-1, 1\}$. It is easy to verify
that, for arbitrary $1 \le i \le n-1$, $\mathcal{O}_z(\alpha, i)$
and $\mathcal{L}(\alpha)$ are contained in the following Alamouti
$$
\begin{pmatrix}
(-1)^{\theta(\alpha,i)}z^*_{\varphi(\alpha, i)} & \phi(\alpha)
z_{\alpha \oplus e_n} \\
(-1)^{\theta(\beta,i)}z^*_{\varphi(\beta, i)} & \phi(\beta) z_{\beta
\oplus e_n},
\end{pmatrix}
$$
where $\alpha \oplus \beta = e \oplus e_i \oplus e_n$. A direct
computation will verify $\varphi(\alpha,i)=\beta\oplus e_n$ and
$\varphi(\beta,i)=\alpha \oplus e_n$. Thus, setting
$\mathcal{L}(\alpha)=\alpha(n+1)\phi(\alpha)z_{\alpha \oplus e_{n}}$
is valid.

Calculating $\theta(\alpha, i) + \theta(\beta, i)$ by definition, we
know that when $i$ is odd, $\theta(\alpha, i) + \theta(\beta, i)
\equiv 0 \pmod 2$; when $i$ is even, $\theta(\alpha, i) +
\theta(\beta, i) \equiv 1 \pmod 2$. Therefore, when $i$ is odd, we
have $\phi(\alpha)=-\phi(\alpha\oplus e_i \oplus e_n)$; when $i$ is
even, we have $\phi(\alpha)=\phi(\alpha\oplus e_i \oplus e_n)$.
Next, we will show a contradiction by calculating the relationship
between
$\phi(\underbrace{0,0,\ldots,0}_{m-1},\underbrace{1,1,\ldots,1}_{m},0)$
and $\phi(1,0,\ldots,1,0)$ in two ways.

\textbf{Way 1:} Let $\alpha = \bigoplus_{i=m}^{2m-1}e_i =
(\underbrace{0,0,\ldots,0}_{m-1},\underbrace{1,1,\ldots,1}_{m},0)$
initially. And let $i=2m-2l+1, l=1,2,\ldots,\frac{m+1}{2}$ and
$i=2l, l=1,2,\ldots, \frac{m-1}{2}$. Finally, we obtain the
relationship between $\phi(\alpha)$ and $\phi(\alpha
\bigoplus_{i=1}^{\frac{m-1}{2}}{(e_{2l}\oplus e_{2m-2l+1})}\oplus
e_m \oplus e \oplus
e_n)=\phi(\bigoplus_{i=1}^{m}{e_{2l-1}})=\phi(1,0,1,0,\ldots,0,1)$.
Since $\phi(\alpha)=-\phi(\alpha\oplus e_i \oplus e)$ if and only if
$i$ is odd, we claim
\begin{equation}
\label{equ_way1}
\phi(\underbrace{0,0,\ldots,0}_{m-1},\underbrace{1,1,\ldots,1}_{m},0)=(-1)^{\frac{m+1}{2}}\phi(1,0,\ldots,1,0).
\end{equation}

\textbf{Way 2:} Let $\alpha = \bigoplus_{i=m}^{2m}e_i =
(\underbrace{0,0,\ldots,0}_{m-1},\underbrace{1,1,\ldots,1}_{m},0)$
initially. And let $i=2m-2l, l=1,2,\ldots,\frac{m-1}{2}$ and
$i=2l-1, l=1,2,\ldots, \frac{m-1}{2}$. Finally, we obtain the
relationship between $\phi(\alpha)$ and $\phi(\alpha
\bigoplus_{i=1}^{\frac{m-1}{2}}{(e_{2l-1}\oplus
e_{2m-2l})})=\phi(1,0,\ldots,1,0)$. Since
$\phi(\alpha)=-\phi(\alpha\oplus e_i \oplus e)$ if and only if $i$
is odd, we claim
\begin{equation}
\label{equ_way2}
\phi(\underbrace{0,0,\ldots,0}_{m-1},\underbrace{1,1,\ldots,1}_{m},0)=(-1)^{\frac{m-1}{2}}\phi(1,0,\ldots,1,0),
\end{equation}
which is contradicted with \eqref{equ_way1}!

Therefore, it's impossible to pad an extra column to
$\mathcal{G}^m_{2m-1}$ to obtain a $[\binom{n}{m-1}, n,
\binom{n-1}{m-1}]$ a COD, and thus COD with parameter
$[\binom{n}{m-1}, n, \binom{n-1}{m-1}]$ does not exist.
\end{proof}

Now, we are ready to prove the final case. Along with Theorem
\ref{thm:Gnw_main}, we have determined the parameters and structures
of all atomic CODs.

\begin{theorem}
\label{thm:Gnm_main} When $n=2m$, let $\mathcal{O}_z[p, n, k]$ be an
atomic COD with some row containing $m+1$ nonzero entries. Then $[p,
n, k] = [\binom{n}{m-1}, n, \binom{n-1}{m-1}]$ or $
[2\binom{n}{m-1}, n, 2\binom{n-1}{m-1}]$.

When $[p,n,k]=[2\binom{n}{m-1}, n, 2\binom{n-1}{m-1}]$,
$\mathcal{O}_z$ is the same as $\mathcal{G}^m_{2m}$ under
equivalence operation. When $[p, n, k] = [\binom{n}{m-1}, n,
\binom{n-1}{m-1}]$, $\mathcal{O}_z$ is the same as
$\mathcal{H}^m_{2m}$ under equivalence operation and $m$ is even.
\end{theorem}
\begin{proof} By deleting the last column of $\mathcal{O}_z[p, n,
k]$, we obtain a COD, say $\mathcal{O}'_z$, with parameter $[p, n-1,
k]$. By Theorem \ref{thm:Gnw_main} and the fact that no column
negation is used, we know $\mathcal{O}'_z$ is the same as the
catenation of $t$ CODs $\mathcal{G}^m_{n-1}$ under equivalence
operation, which are denoted by $\mathcal{O}^{(1)}_z,
\mathcal{O}^{(2)}_z, \ldots, \mathcal{O}^{(t)}_z$. And we denote the
last column by by $\mathcal{L}^{(1)}_z, \mathcal{L}^{(2)}_z, \ldots,
\mathcal{L}^{(t)}_z$, where $\mathcal{L}^{(i)}_z$ and
$\mathcal{O}^{(i)}_z$ are in the same rows of $\mathcal{O}_z$. For
the convenience of description, we denote the row of $\mathcal{O}_z$
by $\alpha^{(i)} \in \mathbb{F}_2^n$, and denote the variable index
of $\mathcal{O}_z$ by $\beta^{(i)} \in \mathbb{F}_2^n,
i=1,2,\ldots,t$.

Let's consider padding the last column. For some variable
$\beta^{(1)}$, recalling Lemma \ref{lem:0pat_to_sc} and the fact
that zero patterns of $\mathcal{G}^w_n$ do not repeat, we know that
it might be in row $\alpha^{(i)}$ of $\mathcal{O}_z$ for some $1\le
i \le t$ and $\alpha$ is uniquely determined.

If $\mathcal{O}_z(\alpha^{(1)},n)=z[\beta^{(1)}]$, then
$z[\beta^{(1)}]$ and other variables $z[\gamma^{(1)}]$ on this row
should be in an Alamouti $2 \times 2$, which implies those
$z[\gamma^{(1)}]$ are in $\mathcal{L}^{(1)}_z$. Repeating this
procedure, we can prove all variables in $\mathcal{O}^{(1)}_z$ are
in $\mathcal{L}^{(1)}_z$, because $\mathcal{O}^{(1)}_z$ is atomic.
Therefore, $(\mathcal{O}^{(1)}_z, \mathcal{L}^{(1)}_z)$ is a COD.
Since $\mathcal{O}_z$ is atomic, we claim $t=1$. By Lemma
\ref{lem:m_odd_notexist}, we know $m$ is even. Since the above
procedure indicates the last column is uniquely determined, we claim
atomic COD with parameter $[\binom{n}{m-1}, n, \binom{n-1}{m-1}]$ is
unique under equivalence operation, which implies $\mathcal{O}_z$ is
equivalent to $\mathcal{H}^m_{2m}$.

If $\mathcal{O}_z(\alpha^{(1)},n)=z[\beta^{(i)}]$ for some $2 \le i
\le t$, without loss of generality, letting $i=2$, then
$z[\beta^{(2)}]$ and other variable $z[\gamma^{(1)}]$ on this row
should be in an Alamouti $2 \times 2$, which implies that those
$z[\gamma^{(1)}]$ are in $\mathcal{L}^{(2)}_z$. Repeating this
procedure, we can prove all variables in $\mathcal{O}^{(1)}_z$
appear in $\mathcal{L}^{(2)}_z$ and all variables in
$\mathcal{O}^{(2)}_z$ appear in $\mathcal{L}^{(1)}_z$, because both
$\mathcal{O}^{(1)}_z$ and $\mathcal{O}^{(2)}_z$ are atomic.
Therefore,
$$
\begin{pmatrix}
\mathcal{O}^{(1)}_z & \mathcal{L}^{(1)}_z \\
\mathcal{O}^{(2)}_z & \mathcal{L}^{(2)}_z
\end{pmatrix}
$$
is a COD.  Since $\mathcal{O}_z$ is atomic, we claim $t=2$. From the
above procedure, we know $\mathcal{L}^{(1)}_z$ and
$\mathcal{L}^{(2)}_z$ are uniquely determined. Therefore, atomic COD
 with parameter $[2\binom{n}{m-1}, n,
2\binom{n-1}{m-1}]$ is unique under equivalence operation, which
implies $\mathcal{O}_z$ is equivalent to $\mathcal{G}^m_{2m}$.
\end{proof}

\section{Conclusion}

\begin{theorem}
\label{thm:poss_para} Given positive integers $p, n, k$, first type COD
$\mathcal{O}_z[p,n,k]$ exists if and only if there exist nonnegative
integers $t_{-1}, t_0, \ldots, t_{\lfloor \frac{n}{2} \rfloor}$ such
that
$$
p = \sum_{i = -1}^{\lfloor \frac{n}{2} \rfloor}{t_i\left( {n \choose
i-1} + {n \choose i+1} \right)} \text{ and } k = \sum_{i=-1}^{\lfloor \frac{n}{2} \rfloor}{t_i {n \choose i}},
$$
when $n \equiv 1, 2, 3 \pmod 4$,
$$
p = \sum_{i = -1}^{\frac{n}{2}-1}{t_i\left( {n \choose i-1} + {n
\choose i+1} \right)} + t_{\frac{n}{2}} {n \choose \frac{n}{2}-1}
$$
and
$$
k = \sum_{i=-1}^{\frac{n}{2} - 1}{t_i {n-1 \choose i-1}} +
t_{\frac{n}{2}} {n-1 \choose \frac{n}{2}-1},
$$
when $n \equiv 0 \pmod 4$.
\end{theorem}
\begin{proof}
\textbf{For the ``if'' direction: } When $n \equiv 1, 2, 3 \pmod 4$,
assume that
$$
p = \sum_{i = -1}^{\lfloor \frac{n}{2} \rfloor}{t_i\left( {n \choose
i-1} + {n \choose i+1} \right)} \text{ and } k = \sum_{i=-1}^{\lfloor \frac{n}{2} \rfloor}{t_i {n \choose i}}. $$ We can construct a COD
achieving parameter $[p,n,k]$ by simply catenating $t_i$ atomic CODs
$\mathcal{G}^i_n[{n \choose i-1} + {n \choose i+1} , n, {n \choose
i}]$, $i=-1,0,\ldots, \lfloor \frac{n}{2} \rfloor$.

When $n \equiv 0 \pmod 4$, assume that
$$
p = \sum_{i = -1}^{\frac{n}{2}-1}{t_i\left( {n \choose i-1} + {n
\choose i+1} \right)} + t_{\frac{n}{2}} {n \choose \frac{n}{2}-1}$$
and $$ k = \sum_{i=-1}^{\frac{n}{2} - 1}{t_i {n-1 \choose i-1}} +
t_{\frac{n}{2}} {n-1 \choose \frac{n}{2}-1}. $$ We can construct a
COD achieving parameter $[p,n,k]$ by simply catenating $t_i$ atomic
CODs $\mathcal{G}^i_n[{n \choose i-1} + {n \choose i+1} , n, {n
\choose i}]$, $i=-1,0,\ldots,  \frac{n}{2} -1$, and
$t_{\frac{n}{2}}$ atomic CODs $\mathcal{H}^{n/2}_n[{n \choose
\frac{n}{2}-1}, n, {n-1 \choose \frac{n}{2}-1}]$.

\textbf{For the ``only if'' direction: } Decompose COD
$\mathcal{O}_z$ into atomic ones. By Theorem \ref{thm:Gnw_main} and
Theorem \ref{thm:Gnm_main}, we know that all atomic CODs have
parameter $[{n \choose i-1} + {n \choose i+1} , n, {n \choose i}]$
for $i=-1,0,\ldots, \lfloor \frac{n}{2} \rfloor$, or $[{n \choose
n/2-1}, n, {n-1 \choose n/2-1}]$ when $n\equiv 0\pmod 4$.

Say, when $n \not\equiv0 \pmod 4$, there are $t_i$ atomic CODs with
parameter $[{n \choose i-1} + {n \choose i+1} , n, {n \choose i}]$, $i=-1,0,\ldots,\lfloor \frac{n}{2} \rfloor$.
When $n \equiv 0 \pmod 4$, there are there are $t_i$ atomic CODs
have parameter $[{n \choose i-1} + {n \choose i+1} , n, {n \choose
i}]$, $i=-1,0,\ldots,n/2-1$, $t''_{n/2}$ atomic CODs with parameter
$[2{n \choose n/2-1}, n, 2{n-1 \choose n/2-1}]$ and $t'_{n/2}$
atomic CODs with parameter $[{n \choose n/2-1}, n, {n-1 \choose
n/2-1}]$. Finally, let $t_{n/2}=2t''_{n/2}+t'_{n/2}$.
\end{proof}

The following corollary characterizes all possible structures of
first type COD, which has similar proof with Theorem
\ref{thm:poss_para}. And thus the proof is omitted.

\begin{corollary} Let $\mathcal{O}_z[p,n,k]$ be a first type COD. Then
$\mathcal{O}_z$ is equivalent to the catenation of $t_{i}$ times
$\mathcal{G}^i_n$, $i = -1, 0, \ldots, \lfloor \frac{n}{2} \rfloor$,
for some $t_{-1}, t_0, \ldots, t_{\lfloor \frac{n}{2} \rfloor}$
satisfying
$$
p = \sum_{i = -1}^{\lfloor \frac{n}{2} \rfloor}{t_i\left( {n \choose
i-1} + {n \choose i+1} \right)}
$$
and
$$
k = \sum_{i=-1}^{\lfloor \frac{n}{2} \rfloor}{t_i {n \choose i}},
$$
when $n \equiv 1, 2, 3 \pmod 4$; $\mathcal{O}_z$ is equivalent to
the catenation of $t_{i}$ times $\mathcal{G}^i_n$, $i = -1, 0,
\ldots, \frac{n}{2}$ and $t'_{\frac{n}{2}}$ times $\mathcal{H}^m_n$,
for some $t_{-1}, t_0, \ldots, t_{ \frac{n}{2}}$ and
$t'_{\frac{n}{2}}$ satisfying
$$
p = \sum_{i = -1}^{\frac{n}{2}}{t_i\left( {n \choose i-1} + {n
\choose i+1} \right)} + t'_{\frac{n}{2}} {n \choose \frac{n}{2}-1}
$$
and
$$
k = \sum_{i=-1}^{\frac{n}{2}}{t_i {n-1 \choose i-1}} +
t'_{\frac{n}{2}} {n-1 \choose \frac{n}{2}-1},
$$
when $n \equiv 0 \pmod 4$.

Furthermore, the number of non-equivalent CODs equals the number of 
different of solutions of $t_{-1}, t_0, \ldots, t_{\lfloor \frac{n}{2} \rfloor}$ when $n \not\equiv 0 \pmod 4$, or $t_{-1}, t_0, \ldots, t_{\frac{n}{2}}, t'_{\frac{n}{2}}$ when $n \equiv 0 \pmod 4$.
\end{corollary}

Since all optimal CODs, which achieves both the maximal rate and
minimal delay, have parameters $[{n \choose m-1}, n, {n-1 \choose
m-1}]$ when $n \equiv 0, 1, 3 \pmod 4$; have parameter $[{n \choose
m-1}, n, {n-1 \choose m-1}]$ when $n \equiv 2 \pmod 4$. And they are
proved to be in first type. We can obtain the following corollary
directly.

\begin{corollary} Let $n = 2m$ or $2m-1$. When $n \equiv 1, 2, 3 \pmod 4$, all maximal-rate, minimal-delay CODs
are the same as $\mathcal{G}^m_n$ under equivalence operation; when
$n \equiv 0 \pmod 4$, all maximal-rate, minimal-delay CODs are the
same as $\mathcal{H}^m_n$ under equivalence operation.
\end{corollary}

The uniqueness under equivalence operation of optimal COD for $n \equiv 0, 1,
3 \pmod 4$ is already proved in \cite{ADK11} by showing that all
such CODs with optimal parameters can be transformed in to a
standard form. The uniqueness for the case $n \equiv 2 \pmod 4$ is
proved for the first time.

In \cite{AKM11}, three facts are proved
\begin{itemize}
\item[1)] For $n=2m-1$, let $\mathcal{O}_z$ be a maximal rate, minimal delay COD.
Then, $\mathcal{O}_z$ is equivalent to a COD that is conjugation-separated.
\item[2)] For $n=2m$, let $\mathcal{O}_z$ be a maximal rate COD
with decoding delay ${2m \choose m-1}$. Then no arrangement
of  $\mathcal{O}_z$ is conjugation-separated.
\item[3)] It is possible to construct a maximum rate COD
with any even number of columns that simultaneously achieves
conjugation-separation and decoding
delay $2{2m \choose m-1}$.
\end{itemize}
By Theorem \ref{thm:Gnw_main}, a $[{2m \choose m-1}, 2m-1, {2m-1 \choose m-1}]$ COD is
equivalent to $\mathcal{G}^m_n$, which conjugation-separated. Thus, 1) is true. By
Theorem \ref{thm:Gnm_main}, we know COD $\mathcal{O}_z[{2m \choose m-1}, 2m, {2m-1 \choose m-1}]$
is equivalent to $\mathcal{H}^m_n$. Therefore, to prove 2), it's sufficient to show
$\mathcal{H}^m_n$ isn't equivalent to a conjugation-separated COD. By the constructions of $\mathcal{G}^m_n$, 3) is true.

\section{Acknowledgment}
We are immensely grateful to Chen Yuan for deciding the signs in
Theorem \ref{thm:Gn}.

\end{document}